\documentclass{article}\usepackage{emulateapj}
\usepackage{psfig}
\usepackage{apjfonts}

\usepackage{epsfig,rotating,amsmath}
\slugcomment{Accepted by \it The Astrophysical Journal\rm}

\received{2002 Sept 6}

\def\gtrapprox{\;\lower 0.5ex\hbox{$\buildrel >\over \sim\ $}}
\def\lessapprox{\;\lower 0.5ex\hbox{$\buildrel < \over \sim\ $}}          

\def\NIIHa {[N${\rm\scriptstyle II}$]/H$\alpha$}
\def\NII   {[N${\rm\scriptstyle II}$]}

\def\deg   {$^\circ$}

\def\HI    {H{$\rm\scriptstyle I$}}
\def\Ha    {${\rm H}\alpha$}
\def\eg    {{\it e.g.\ }}

\def\etal  {{\it\ et al.}}
\def\kms   {\ km s$^{-1}$}
\def\intensity{\ifmmode{{\rm erg\ cm}^{-2}{\rm\ s}^{-1}
      {\rm\ Hz}^{-1}{\rm\ sr}^{-1}}
      \else {erg cm$^{-2}$ s$^{-1}$ Hz$^{-1}$ sr$^{-1}$}\fi}
\def\Em{\ifmmode{{\cal E}_m}\else {{\cal E}$_m$}\fi}
\def\Dm{\ifmmode{{\cal D}_m}\else {{\cal D}$_m$}\fi}
\def\fesc{\ifmmode{\hat{f}_{\rm esc}}\else {$\hat{f}_{\rm esc}$}\fi}
\def\fescs{\ifmmode{f_{\rm esc}}\else {$f_{\rm esc}$}\fi}
\def\tauLL {\ifmmode{\tau_{\scriptscriptstyle LL}}\else 
           {$\tau_{\scriptscriptstyle LL}$}\fi}

\begin{document}
\title{\Ha\ Emission from High-Velocity Clouds and their Distances}

\author{M. E. Putman\altaffilmark{1,2},
        J. Bland-Hawthorn\altaffilmark{3},
        S. Veilleux\altaffilmark{4,5,6},
        B. K. Gibson\altaffilmark{7},
        K. C. Freeman\altaffilmark{8},
	P.R. Maloney\altaffilmark{1}}

\altaffiltext{1}{Center for Astrophysics and Space Astronomy, University of Colorado, Boulder, CO 80309-0389; mputman@casa.colorado.edu, maloney@origins.Colorado.edu} 
\altaffiltext{2}{Hubble Fellow}
\altaffiltext{3}{Anglo-Australian Observatory, P.O. Box 296, Epping, NSW 2121, Australia; jbh@aaoepp.aao.gov.au}
\altaffiltext{4}{Dept. of Astronomy, University of Maryland, College Park, MD 20742; veilleux@astro.umd.edu}
\altaffiltext{5}{Cottrell Scholar of the Research Corporation}
\altaffiltext{6}{Current address: 320-47 Downs Lab., Caltech, Pasadena, CA 91125 and
Observatories of the Carnegie Institution of Washington, 813 Santa
Barbara Street, Pasadena, CA 91101; veilleux@ulirg.caltech.edu}
\altaffiltext{7}{Centre for Astrophysics \& Supercomputing,
                 Swinburne University, Mail \#31, P.O. Box 218,
                 Hawthorn, VIC, Australia 3122; bgibson@astro.swin.edu.au}
\altaffiltext{8}{Research School of Astronomy \& Astrophysics,
                 Australian National University, Weston Creek P.O.,
                 Weston, ACT, Australia 2611; kcf@mso.anu.edu.au}

\begin{abstract}
We present deep \Ha\ spectroscopy towards several high-velocity clouds
(HVCs) which vary in structure from compact (CHVCs) to the Magellanic
Stream.  The clouds range from being bright ($\sim$640 mR) to having
upper limits on the order of 30 to 70 mR.  The \Ha\ measurements are
discussed in relation to their \HI\ properties and distance constraints
are given to each of the complexes based on \fesc\ $\approx$ 6\% of the
ionizing photons escaping normal to the Galactic disk (\fescs\ $\approx 1 - 2$\%
when averaged over solid angle).  
The results suggest that many HVCs
and CHVCs are within a $\sim$40 kpc radius from the Galaxy and are not
members of the Local Group at megaparsec distances.  However, the Magellanic Stream
is inconsistent with this model and needs to be explained.  It has 
bright \Ha\ emission and little 
[NII] emission and appears to fall into a different category than the 
currently detected HVCs.  This may
reflect the lower metallicities of the Magellanic Clouds compared to
the Galaxy, but the strength of the \Ha\ emission cannot be
explained solely by photoionization from the Galaxy.  The 
interaction of the Stream with halo gas or the presence of yet unassociated young
stars may assist in ionizing the Stream.
\end{abstract}


\section{Introduction}

The smooth accretion of gas onto galaxies allows for continuous galaxy
evolution and star formation.  The intergalactic gas which feeds
galaxies is seen in absorption against a bright background source along
filaments of galaxies (e.g. Penton, Stocke \& Shull 2002) and is
predicted by simulations of the ``cosmic web'' (e.g. Dav\'e\etal\ 1999).
When this gas reaches a certain radius from the galaxy, it may be able
to condense and cool, and in the case of our own Galaxy, the gas could
become observable in 21-cm emission.  Together with the remnants of
Galactic satellites, these objects may be represented by the
high-velocity clouds (Oort 1966).

High-velocity clouds are concentrations of neutral hydrogen which do
not fit into a simple model of Galactic rotation and cover 30-40\% of
the sky (e.g.~Wakker \& van Woerden 1991; Lockman et al. 2002).  There have
been several models which propose that HVCs are the primordial building
blocks of galaxies, the leftovers along the supergalactic filaments.
Blitz\etal~(1999) and Braun \& Burton (1999) proposed HVCs, in
particular the compact HVCs (CHVCs), represent the missing satellites 
of the Local Group, at mean distances of $\sim$1 Mpc.  
These models have been called 
into question (e.g. Zwaan 2002; Sternberg, McKee \& Wolfire 2002; Maloney \& Putman 2003).

\Ha\ observations provide a direct test of whether HVCs are infalling
members of the Local Group at large distances from the Galaxy.  Models
of the Galactic ionizing radiation field indicate that ionizing photons
are capable of reaching distances on the order of 100 kpc; HVCs
can act as an \HI\ screen and the \Ha\ emission measure reflects the
ionizing photon flux reaching the cloud
(Bland-Hawthorn \& Maloney 1999, hereafter B99; Bland-Hawthorn \& Maloney
2002, hereafter B02).  This is confirmed by recent
\Ha\ observations of large high velocity complexes which
have direct distance bounds of $<$ 10 kpc (Tufte\etal\ 1998; hereafter T98).  If any of the HVCs are at distances on the order
of 1 Mpc they should not be detectable, as the cosmic ionizing background
is too low; therefore, any detection of
\Ha\ emission brings the HVCs within the extended Galactic Halo.
\Ha\ observations of HVCs with known distances also provide insight
into how the ionizing radiation escapes from the Galactic disk, other
ionization processes present in the Galactic halo, and the nature of the halo/IGM
interface. 

In this paper we present HVC optical line emission observations to
investigate the relationship between HVCs and the Galaxy.  The paper
begins by summarizing the Fabry-Perot and long slit \Ha\ observations
in $\S$2, and presents the results of the observations in $\S$3. 
In $\S$4-5, we
discuss our findings and interpret them in the context of the location
and environment of the HVCs.
The ionization of the Magellanic Stream is considered in $\S$6 and an overview
of the results is presented in $\S$7.

\section{Observations}

The Fabry-Perot \Ha\ observations were obtained in five Anglo-Australian Telescope (AAT) observing runs from
December 1997 - June 1999 and one William Herschel Telescope (WHT) run in January 1999.  At both
sites, the TAURUS-2 interferometer was used in conjunction with the
University of Maryland 44$\mu$m etalon.  Single orders of interference
were isolated using 4-cavity blocking filters with high throughput
(80$-$90\%) and bandpasses well matched to the etalon free spectral
ranges.  The focal plane was baffled to give either a 10\arcmin\ or,
on the WHT (northern objects), a 5.0\arcmin\ field.  The resulting 
ring pattern covered about 45{\AA} (\Ha, \NII$\lambda$6583) at
the AAT and
20{\AA} (\Ha) at the WHT.  The resolution is 1{\AA} (or 46\kms~at \Ha).
The repeated exposures were generally 10$-$20 minutes.  A deep sky
exposure was also made in a region 5\deg$-$20\deg\ away from the cloud,
at a position that does not contain any high-velocity \HI\ based on
HIPASS limits ($< 2 \times 10^{18}$ cm$^{-2}$).  The reduction and
analysis is discussed in Bland-Hawthorn\etal\ (1998; hereafter B98).

HVCs and corresponding deep sky exposures were also observed with the Double Beam Spectrograph (DBS) on the
Siding Spring Observatory 2.3 m telescope over 5
observing runs from August 1998 - April 2000.  The DBS was set up with
a 7$^{\prime}$ long slit and a slit width of 2$^{\prime\prime}$,
yielding a spectral resolution of 0.57 A (26 \kms~at \Ha).  
Spectral reduction was done using the VISTA and IRAF reduction packages.
The two-dimensional spectra
from the large sets of exposures obtained on target and sky
were each reduced separately. The procedure involved bias subtraction
using both bias frames and the overscan area on each exposure,
flatfielding using QI lamp
exposures, and cosmic ray removal with a 2.5-$\sigma$ high/low filter.
S-distortions and illumination effects along the slit were removed at
this stage using the positions and intensity profiles of the sky
lines. 
The central 240 rows 
of each 2D spectrum were extracted and then
wavelength calibrated based on the positions of the sky lines in the
object's spectrum (only possible with the red spectra) and the NeAr
lamp spectra obtained at the same airmass as, and immediately before
or after, the object's exposure. The spectra were put on an absolute flux scale using the spectra of flux
standards obtained throughout the night.
Each objects' exposures were added together after aligning the spectra using the closest skyline to the
expected emission from the HVC as a guide.  An example of a DBS spectrum is shown
in Figure 1.

We examined the deep sky exposures closely for signs of \Ha\ emission
at a velocity similar to the closest detectable \HI\, and found no
indication of emission.  This was especially important to check
considering the OVI that has been detected in absorption at the velocities
of nearby \HI\ HVCs, but off the \HI\ contours of these clouds (Sembach et al. 2003),
and the extended low \HI\ column density emission found around cataloged HVCs (Lockman et al. 2002).
We assume foreground Galactic extinction
along a given sight line, measured from the COBE/DIRBE maps
(Schlegel\etal\ 1998), and therefore correct all \Ha\ emission measures (E$_m$)
for dust extinction.  We also include the uncorrected E$_m$ values
in Tables 1 and 2, as the dust correction may not be realistic for
the low latitude clouds.  E$_m$ upper limits quoted throughout this
paper are 2-sigma for the 
TAURUS data and 3-sigma for the DBS data.  Our characteristic detection errors
are approximately 10 mR if the \Ha\ detection is at least 2{\AA} from a skyline, 
but close to a skyline the errors can reach 15-30 mR.
Future use of the nod+shuffle technique (Glazebrook \& Bland-Hawthorn 2001) with 
the Fabry-Perot staring method may be able to reach levels of $\sim$5 mR.

Many of the observed HVCs were first identified by
HIPASS (see Putman\etal\ 2002a; hereafter P02).  This is
especially true of the CHVCs.  The clouds observed were chosen
because they have:  \HI\ velocities which isolate an equivalent velocity
\Ha\ line from the skylines, a proposed extragalactic nature (e.g. the
CHVCs), and/or an estimate has been made of their distance and/or
origin (e.g. Complex M, Magellanic Stream).  In the latter case, the
observations could be used to clarify the nature of the \Ha\ emission.
Several positions observed by Weiner \& Williams (1996; hereafter WW96) and T98 were
repeated to compare observing and reduction methods.

\section{Results}

The results of the observations are described in Table 1 and 2.  Table 1
lists the positive detections and Table 2 the non-detections.  The objects
are grouped in terms of their high-velocity classification and are
named either by their traditional name or by
their P02 classification, which is the type of cloud (CHVC = Compact HVC, :HVC =
slightly more extended than a CHVC, HVC = extended HVC, or XHVC = a HVC which has
\HI\ emission that merges with Galactic velocities), followed by the intensity weighted Galactic
longitude and latitude and the central LSR velocity.  The \HI\ properties are from P02 (excluding 
the northern targets which are from the LDS (e.g. Complexes H and M)) and 
are always taken along the sightline of the \Ha\ observation. 
The results of B98, T98, and Tufte et al. (2002; hereafter T02) are also included in Table 1. 
The columns of Table 1 are:  $\ell$ and $b$
coordinates of the \Ha\ observation, HVC name, \HI\ column density, \HI\ velocity (LSR),
\HI\ velocity width, the extinction corrected \Ha\ emission measure with W or D 
in parentheses if the result is from WHAM or the DBS respectively, the value of the \Ha\ emission measure before the
extinction correction, the [NII]$\lambda$6583/\Ha\ ratio, the velocity of the \Ha\ detection (LSR), and
the predicted distance to the HVC based on its $\ell$, $b$, and extinction corrected
emission measure (see $\S$4).   Some of the [NII]/\Ha\ ratios are not included
due to the observation not including the wavelength of the [NII]$\lambda$6583 line (i.e.
the WHT and WHAM observations).  Table 2 does not include the [NII]/\Ha\ ratio or the predicted distance
(see $\S$5), but does include two limits on [OIII] emission.

The close relationship between \HI\ velocity and \Ha\ velocity is shown in Figure
2 and
the complete lack of correlation between the \Ha\ emission measure and
\HI\ column density is shown in Figure 3.  This is what would be expected if the outer skin of
the HVC is being ionized by an external ionizing radiation field. 
Figure 2 also shows that non-detections (open diamonds) span the entire range of high velocities.
Though not shown, there is also no relationship between the strength of the \Ha\
emission and the velocity of the HVC (in the LSR or GSR reference frame).
Figure 3 shows that undetected clouds span the entire range of \HI\ column densities, i.e. there does
not currently seem to be a lower or upper column density cutoff.  The
distribution of the \Ha\ detections and non-detections on the sky in
Galactic coordinates is shown in Figure 4, and a large number of the
\Ha\ observations are depicted on the \HI\ map of the Magellanic System shown
in Figure 5.  We now
discuss the specific detections listed in Table 1 and the undetected clouds listed
in Table 2.  Pictures and spectra of most of the high-velocity complexes are shown in Putman (2000).

\subsection{Detections}

{\bf Complexes:}
Several of the HVCs detected in \Ha\ are part of larger complexes which are defined
by Wakker \& van Woerden (1991).  The \Ha\ brightest of these is
Complex L, a negative velocity HVC made up of several clumpy 
filaments, with several small clouds scattered amongst the 
filaments.  The cloud mapped here is HVC341.6+31.4-142 in the
P02 catalog and the brightest emission lies closest to the head of the cloud.
Complex L has a highly elevated [NII]/\Ha\ ratio (2.7).
Along with the detections there was one non-detection in a very low column
density ($\sim 10^{18}$ cm$^{-2}$) part of Complex L.
All of the positions with bright detections have
column densities between 1.6 - 3.6 $\times 10^{19}$ cm$^{-2}$.

The other detected complexes are:  Complex M, which has an upper distance
constraint of $<$ 4 kpc (Ryans et al. 1997) and was detected at a similar level
by T98, Complex H,  which lies along the Galactic
Plane making this detection more tentative (especially since the \Ha\
velocity is offset by 30 \kms\ from the \HI\ velocity), and Complex GCP
(Smith Cloud) which was originally presented in B98 and now has a limit
on the [OIII] emission at the position of Smith1 ($<$ 70 mR; Table 2).
We include the T98 detections of Complexes A and C with a model distance
because they have direct distance limits of 4-10 kpc (van Woerden et al. 1999)
and $>$ 6 kpc (Wakker 2001), respectively.

{\bf The Magellanic Stream:}
 The Magellanic Stream shown in Figure 5 is the result of the interaction of the Large
and Small Magellanic Cloud with each other and the Galaxy.  It trails
the Magellanic Clouds for over 100$^{\circ}$ through the South
Galactic Pole and has a velocity gradient of 700 \kms~from head to tail (relative
to the Local Standard of Rest; 400 \kms~relative to the Galaxy).
The Stream is a complicated network of
filaments and clumps, but remains relatively continuous along its
entire length (see Putman et al. 2003; hereafter P03).  Stars have not yet been
found in the Stream (e.g. Guhathakurta \& Reitzel 1998), but  \Ha\
emission has been previously detected  by WW96 at the 
level of 200 - 400 mR.

We observed several positions
along the Magellanic Stream, including one repeat of a WW96
observation, with both TAURUS and the DBS.  The repeat observation of
MSIIa is approximately the same as WW96 with TAURUS, but is lower
with the DBS.  This could be due to the difference in the field of view of
TAURUS and the DBS (a 10\arcmin\ diameter FOV versus a 7\arcmin\ $\times$ 2$^{\prime\prime}$ slit).  [NII] was also detected and the ratio to \Ha\ is
low compared to the Smith Cloud and Complex L (0.15 vs. 0.6 - 2.7).  MSIIa was
subsequently observed in [OIII]$\lambda$5007 and no detection was
obtained ($<$ 52 mR; Table 2).  As tabulated in Table 1, a new relatively weak
\Ha\ detection was made at the head of the Stream, ($\ell, b$) = 304\deg, -67\deg, and at the position
of the background QSO Fairall~9 where OVI absorption has also been detected
($10^{14.3}$ cm$^{-2}$; Sembach et al. 2003). There was also a non-detection
at ($\ell, b$) = 293.4\deg, -56.4\deg\ and at the 
tail of the Stream (MSV; ($\ell, b$) = 96.5\deg, -53.9\deg) as tabulated in
Table 2.  
As shown in Figure 5, there are large variations in the strength of the \Ha\ emission along
the Stream's length, and so far there does not seem to be a correlation
with the \HI\ column density (Figure 3).   However, one should consider that the beam used
in the \HI\ observations is larger than the FOV of TAURUS (15.5\arcmin\ vs. 10\arcmin).
Though the number of observations remains limited, there also does not seem to be
a gradient of \Ha\ brightness along the Stream.  Currently, the brightest
detection is approximately at the South Galactic Pole in a region of complexity
in terms of the high-velocity \HI\ gas distribution (P03).   The velocities of the \HI\ and \Ha\ lines
generally closely agree (within $\sim$10 \kms; Figure 2).
Several positions along the Magellanic Bridge and Leading Arm were also observed.  All
of the  pointings were non-detections (see Table 2 and Figure 5), except for an extremely bright
observation at the position of a known OB association (Bridge M in
Table 1; see also Marcelin\etal\ (1985)).

{\bf Compact High-Velocity Clouds (CHVCs):}
Two compact high-velocity clouds (CHVCs) were detected with the DBS.
CHVC197.0-81.8-184 is located $\sim10$\deg\ from the Stream where it
passes through the South Galactic Pole (see Figure 5).  The \Ha\ detection of this cloud (Fig. 1) is at
the level of many of the Stream detections.  
 The second CHVC is a very small and isolated cloud located
in the region leading the LMC (Figure 5).  CHVC266.0-18.7+336 has a velocity which places the \Ha\ line
at the edge of a skyline, making the brightness of this detection 
somewhat less certain.  The CHVC detections of T02 with model distances are
also included in Table 1.

\subsection{Non-Detections}

There are several clouds which were not detected in
this survey and are summarized in Table 2.  Some of these clouds have detections
reported in the conference proceedings of Weiner et al. (2001), but the
precise coordinates of their observations have not yet been reported.  
This is not unusual considering
the range of detections and non-detections noted in the previous section within the same high-velocity complex. 
Many of the HVCs which we have only non-detections for lie in 
approximately the same region of the sky (see Figure 4).  
The undetected clouds mostly lie in the Galactic Longitude range of $\ell = 250$\deg $-$ 320\deg, and include the length of the Leading Arm of the Magellanic
System (Figure 5), several HVCs and CHVCs, and part of the Extreme Positive Velocity Complex.
We note that many of these clouds (marked with a $*$ in Table 2) have velocities that place the
\Ha\ line close to a skyline, making the non-detections somewhat less certain.  

Additional non-detections include the high positive velocity cloud HIPASS J1712-64 (Kilborn\etal\ 2000) which has an
\Ha\ upper limit of 44 mR, and the clouds associated with the Sculptor
dSph galaxy by Carignan\etal~(1998) (cataloged as CHVC286.3-83.5+091 and
CHVC290.6-82.8+095 in P02).  It is unclear if these clouds are
actually associated with the Sculptor dSph.  The \HI\ maps of P03
and Carignan (1999) show the complexity of this region in high-velocity gas, with
a high concentration of clouds at similar and very different velocities to the Sculptor dSph.
There is an undetected negative velocity XHVC at approximately -145 \kms~along
our observed sightline to the clouds associated with the Sculptor dSph, as well as a nearby positive velocity XHVC, which was also undetected.

\section{The \Ha\ Distance Constraint}

The \Ha\ distance constraint is based on photoionizing radiation
escaping from the Galactic disk and ionizing the surface of \HI\
clouds within the Galactic halo (B98).  It relies on our knowing the 
strength and morphology of the halo ionizing field, and can be
affected by a cloud's covering fraction,
topology, and orientation to our line of sight (B02).  Variations in \Ha\
brightness across a single HVC may be due to these issues, and we stress that
the \Ha\ brightest point on the HVC (i.e. the point on the cloud receiving the
most ionizing photons from our Galaxy) is the measure that should be used when
estimating the HVC distance.  Since we will not know if we have observed the brightest
point on a particular HVC until we are able to do large scale \Ha\ mapping of each cloud,
our far field distance estimates in Table 1 currently serve as upper limits.
 Several HVCs 
with strong direct distance contraints (see Wakker (2000) for a
summary) have now been detected
in \Ha\ by WHAM (T98), Weiner\etal\ (2001),
and this survey.  There is also an IVC (Complex K; Haffner\etal\ 2001) that 
has been completely mapped in \Ha\ emission and has a distance constraint.  
The \Ha\ emission measures from these clouds are consistent with the 
model predictions of B99 $-$ updated in B02 to include spiral arms $-$ 
which uses an escape fraction normal to the disk of \fesc\ $=$ 6\% 
(\fescs\ $\approx$ $1-2$\% averaged over $4\pi$ sr).  The escape fraction
used in the B02 spiral arm model has been adopted based on its agreement with the direct distance 
determinations and \Ha\ emission measures for Complex A, M, C, and
the IVC, Complex K.  It has a factor of two uncertainty which could 
affect the predicted distances listed in Table 1 by 50\%.
Figure 6 shows the effect of using a model with spiral arms compared
to exponential and uniform disk models.  The halo ionization field
is very different for a dusty spiral versus an exponential disk within
10 kpc of the Galactic disk.

All of the HVCs detected in \Ha\ emission would be at distances
within 40 kpc in the context of this model.  The detection of two CHVCs indicates that
some fraction of this population falls within the extended Galactic halo.
This is supported by the CHVC detections of T02.  These CHVCs
would be within $\sim$13 kpc using this distance determination 
method.  The model prediction for a radius vector towards Complex L is shown in
Fig. 7. Note that the spiral
arm model predicts that Complex L lies directly over 
a spiral arm, but there is a near and far field solution, depending
on its exact position.
There is some indication that HVCs along sightlines over spiral arms
are brighter, as expected for clouds within about 10 kpc (B02), but 
more sightlines are needed to confirm this.

Though the detection of \Ha\ emission argues for HVCs being within
the Galactic halo, the brightness of the Magellanic Stream detections
needs to be understood before the distance constraint can be considered 
fully reliable (see $\S$6 and Bland-Hawthorn \& Putman 2001, hereafter B01).  
We also note that Complex L and GCP (the Smith Cloud) 
not only have high \Ha\ emission measures (which makes sense, as they 
most likely lie inside the solar circle above the spiral arms), but also 
elevated \NIIHa\ emission.  The [NII] emission may be an indication 
of enhanced electron temperatures (Reynolds, Haffner \& Tufte 1999), 
rather than the presence of an alternative source of ionization (\eg shocks). 
There are a variety of ways to produce this effect (\eg photoelectric heating
(Wolfire\etal\ 1995)), and the enhanced low-ionization
emission is also seen in the high latitude gas of spirals 
(Haffner\etal\ 1999; Veilleux\etal\ 1995; Miller \& Veilleux
2003a, 2003b). In
essence, we can use the elevated \NIIHa\ to argue that some HVCs are more
than several kiloparsecs from the plane, and comprise part of the extended
ionized atmosphere seen in external galaxies.  Further support comes from 
\HI\ structure of these clouds, each of which show possible extensions
into Galactic \HI.

\section{Do non-detections correspond to large distances?}

If the \Ha\ normalization to local HVCs is valid, this may indicate
that some HVCs which are faint or undetected in \Ha, particularly those
at high latitude, are dispersed throughout the extended halo on scales
of 50~kpc or more.  The cosmic
ionizing background radiation ($\sim 10^4$ phot/s; Maloney \& Bland-Hawthorn 1999) 
would correspond to a 5 mR \Ha\ detection and would only begin to dominate
over the Galactic ionizing radiation field approximately 100 kpc from
our Galaxy. 
Considering the \Ha\ upper limits in some cases and the variations
in intensity across the HVCs, it
remains to be seen whether most of the clouds which have non-detections
are actually at large distances from the Galactic Plane.  \Ha\ mapping 
across an entire HVC to find the brightest
\Ha\ emission, higher
resolution \HI\ observations to clarify the column density at
the position of the \Ha\ observation, and the development of models of the
escape of ionizing radiation from the Galactic Plane will help resolve the
non-detection issue.  
It may be that some clouds will remain undetected in 
certain directions if they lie at too low an angle from our viewpoint, or 
do not lie above spiral arms or HII regions.  Shadowing
and the size of the TAURUS beam may also be important considerations.  There may be an
{\it observed} relationship between the strength of \Em\ and the
position of the cloud above the Galaxy, as clouds at $\ell >$ 330\deg\
and $\ell <$ 60\deg\ have a slight tendency to be brighter and clouds
between $\ell = 250 - 320$ remain largely undetected (Figure 4).  This is expected
from their line of sight over the Galaxy (see Taylor \& Cordes 1993) 
and from the B02 model.

\section{What is ionizing the Magellanic Stream?}

The Stream is brightest at the South Galactic Pole and fainter towards
the head and tail.  This would be expected for halo gas ionized by 
an opaque disk where ionizing photons escape preferentially along the
Galactic poles (B99).  The match
between the \HI\ velocity and the \Ha\ velocity for all clouds supports
photoionization.  However, if ionizing photons from the Galaxy are
reaching HVCs at distances of $\sim$10 kpc, why are Stream positions
near the South Galactic Pole, which most likely lie at distances
between 20 $-$ 100 kpc (Gardiner 1999; Moore \& Davis 1994),
consistently brighter than the HVCs?   As shown in Figure 8, 
at a mean Stream distance of 55~kpc, the expected emission measure 
of a flat \HI\ stream is 30$-$50~mR (B02), an order of magnitude 
fainter than the brightest detections.  Figure 8 also shows that the
contribution from the LMC will not play a dominant role in ionizing
the majority of the Stream.

Is it possible that sections of the Stream are just that much closer to
the Galaxy disk than the Magellanic Clouds?  With the detection of the
head of the Stream (Fairall~9 sightline), this possibility seems
unlikely, as the head of the Stream is presumed to be close to the
Magellanic Clouds (50-60 kpc).  Thus the distances predicted in Table 1
for the Stream sightlines are not relevant and we need to look for
another source of ionization in the Stream.  The detection of O~VI
absorption in and around the Stream may provide some clues
(Sembach\etal\ 2003).  Interaction with a halo medium could provide
some pre-ionization which could elevate the Stream's \Ha.  The outer
halo medium may well be clumpy, particularly at the poles, from the
leftovers of other satellites or from self-interaction of the Stream
(B01; P03).  CHVC197.0-81.8-184 may represent some of this
debris.  This CHVC is only 10\deg\ from the main filament of the Stream
and is as \Ha\ bright as the Stream, possibly indicating a large spread
of debris associated with Stream's \Ha\ emission.  Two of the T02 detected
CHVCs (shown in Fig. 5) may also represent the spread of ionized Stream debris.

Another possibility is that there are stars associated with the Stream which have
yet to be detected.  Recent results have found small isolated HII regions in interacting
systems that can be ionized
by a few O stars (e.g., Gerhard et al. 2002; Ryan-Weber et al. 2003).  This indicates that isolated star 
formation can be triggered in low
density interactive debris, which could in turn play an important role in ionizing this material.  A single massive
O star 1 kpc from the Stream could lead to an emission measure of 40 mR.  If
the star was actually embedded in the Stream this contribution would
obviously be much higher.  White dwarfs would not significantly
contribute to the ionization of the Magellanic Stream unless their
density was much higher than that found in the solar 
neighborhood (Bland-Hawthorn, Freeman, \& Quinn 1997).
Thus far, only limited areas of the Stream have been surveyed for stars.
Ongoing and future stellar surveys will provide further insight into
the possibility of the Stream harboring young, ionizing stars.  

\section{Overview}

The \Ha\ observations presented here are a combination of detections
and non-detections on clouds with \HI\ column densities greater than
a few times 10$^{18}$ cm$^{-2}$.  This represents the complex nature of the ionized
component of HVCs and the importance of mapping across an entire cloud
before accepting a non-detection as meaningful for the entire
high-velocity complex.  The results thus far show a population of
clouds which appear to extend out of Galactic \HI\ emission, are
\Ha\ bright, and show an elevated \NIIHa\ ratio, as well as an
undetected population which tend to be in a specific region of Galactic
longitude and are relatively isolated from Galactic emission.  The
detection of several CHVCs in both this paper and the T02 paper indicates that many of these clouds are indeed within the
Galactic halo.  The non-detections of
some CHVCs cannot be used to argue for a greater distance until the
origin of the non-detections in other complexes is understood.  

The \Ha\ emission measures of the clouds with distance constraints are
consistent with the surfaces of the clouds being ionized by $\sim 6$\%
of the Galaxy's ionizing photons.  All of the clouds detected here are within
40 kpc of our Galaxy based on their level of \Ha\ emission.
 The Magellanic Stream appears to fall into a different category than
the currently detected HVCs, with bright \Ha\ emission but little or no
[NII] emission, possibly due to the lower metallicities of the
Magellanic Clouds compared to the Galaxy.  The strength of the
\Ha\ emission cannot be easily explained by photoionization from the
Galaxy alone, and it is possible that interaction with halo debris, or the
presence of yet unassociated young stars, is partially
responsible for the Stream's elevated \Ha\ emission.
 Through future \Ha\ observations which include mapping head-tail \HI\
clouds, the length of the Magellanic Stream, OVI absorption sightlines, and complexes of known
distance, and the development of models which trace the path of the
escaping photons from the Galactic Plane, we may come to a consensus on the
origin of the \Ha\ emission in all high-velocity clouds.

\acknowledgments{We thank our referees, Robert Benjamin and an anonymous reviewer, for 
many useful comments, and Ben Weiner and Mike Shull 
for helpful discussions.  M.E.P. acknowledges support by
NASA through Hubble Fellowship grant HST-HF-01132.01 awarded by the Space Telescope Science
Institute, which is operated by AURA Inc. under NASA
contract NAS 5-26555.
S.V. is indebted to the California Institute of Technology and the
Observatories of the Carnegie Institution of Washington for their
hospitality, and is grateful for partial support of this research by a
Cottrell Scholarship awarded by the Research Corporation, NASA/LTSA
grant NAG 56547, and NSF/CAREER grant AST-9874973.
 B.K.G. acknowledges the support of the Australian Research
Council, through its Large Research Grant and Discovery Project schemes.}

\clearpage

\centerline{\psfig{figure=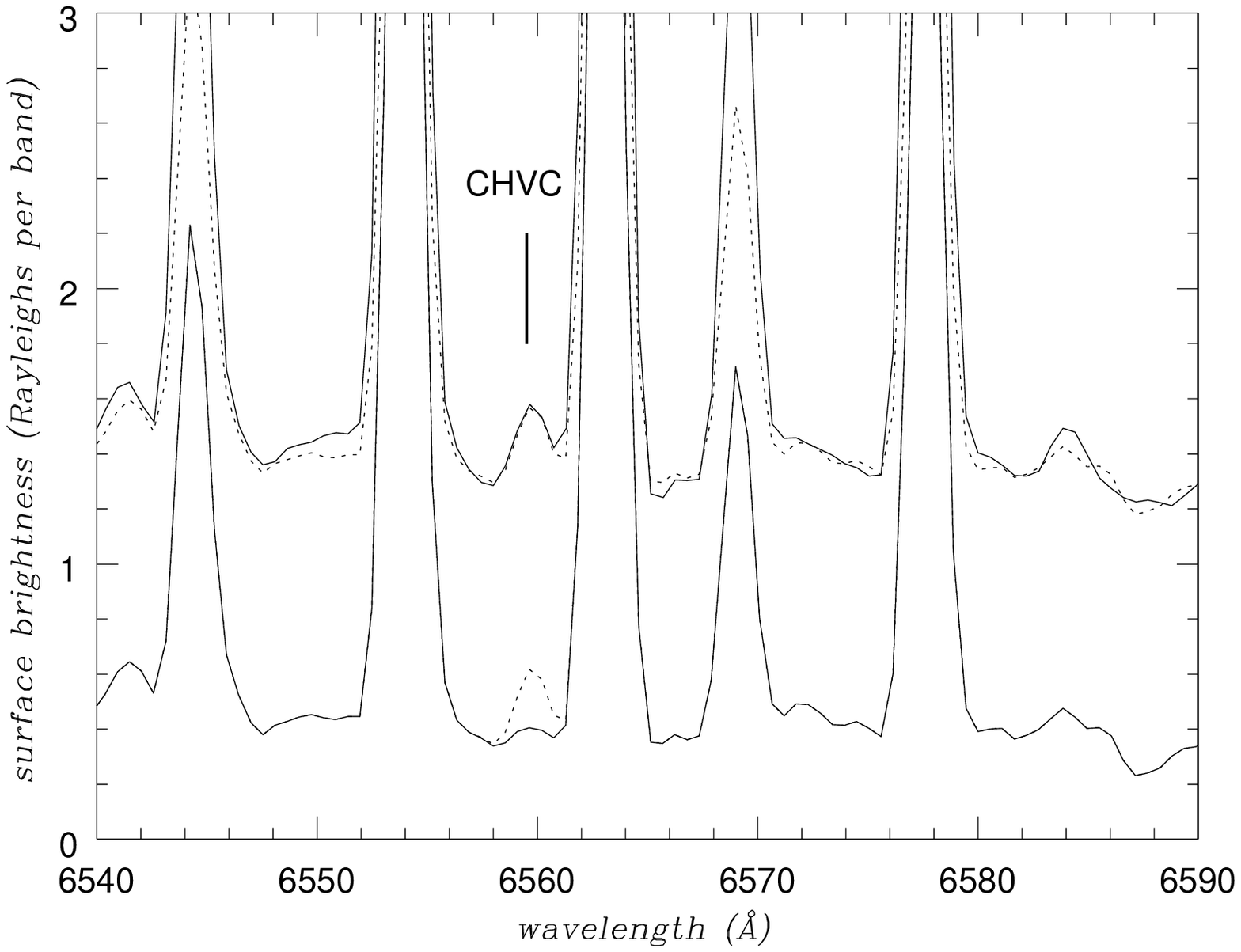,width=8cm,angle=0}}

\noindent {\sl Fig.~1.} DBS spectrum of CHVC197.0-81.8-184 showing
\Ha\ emission at the level of 220 mR.  The top spectrum is the CHVC
observation (solid line) with the sky observation with a gaussian fit
at the velocity and \Ha\ strength of the CHVC overplotted (dashed
line).  The bottom plot shows the sky spectrum with the gaussian fit to
the \Ha\ detection shown as the dashed line.
\medskip

\centerline{\psfig{figure=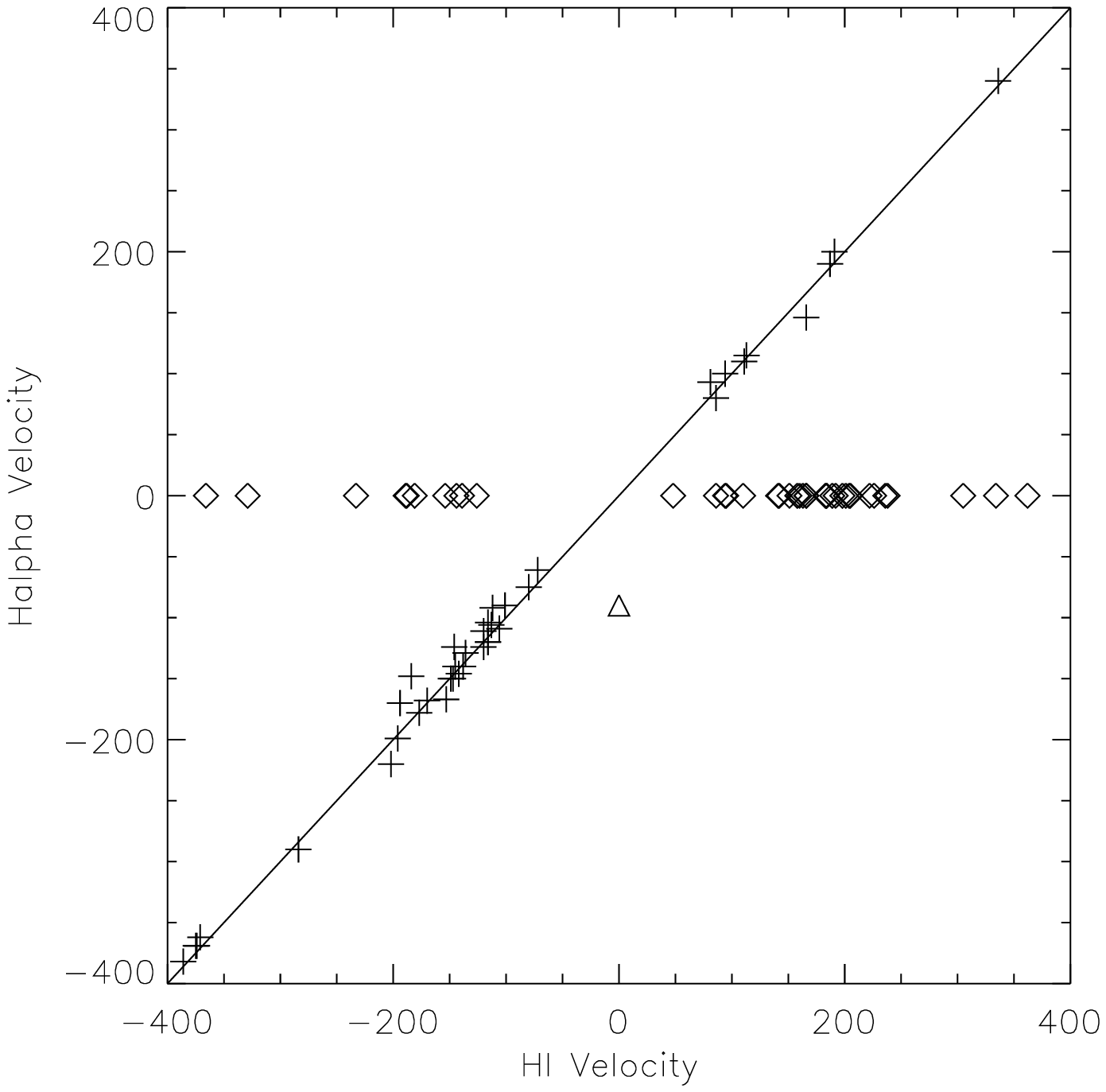,width=10cm,angle=0}}

\noindent {\sl Fig.~2.}
The relationship between the \Ha\ and \HI\ velocity for all of the recently published
HVC \Ha\ observations (this paper; WW96; T98; B98; T02).  Crosses show the detections,  diamonds the non-detections in \Ha, and the triangle is the one high-velocity detection in \Ha\ but not \HI\ on the edge of Complex M (T98).

\medskip

\centerline{\psfig{figure=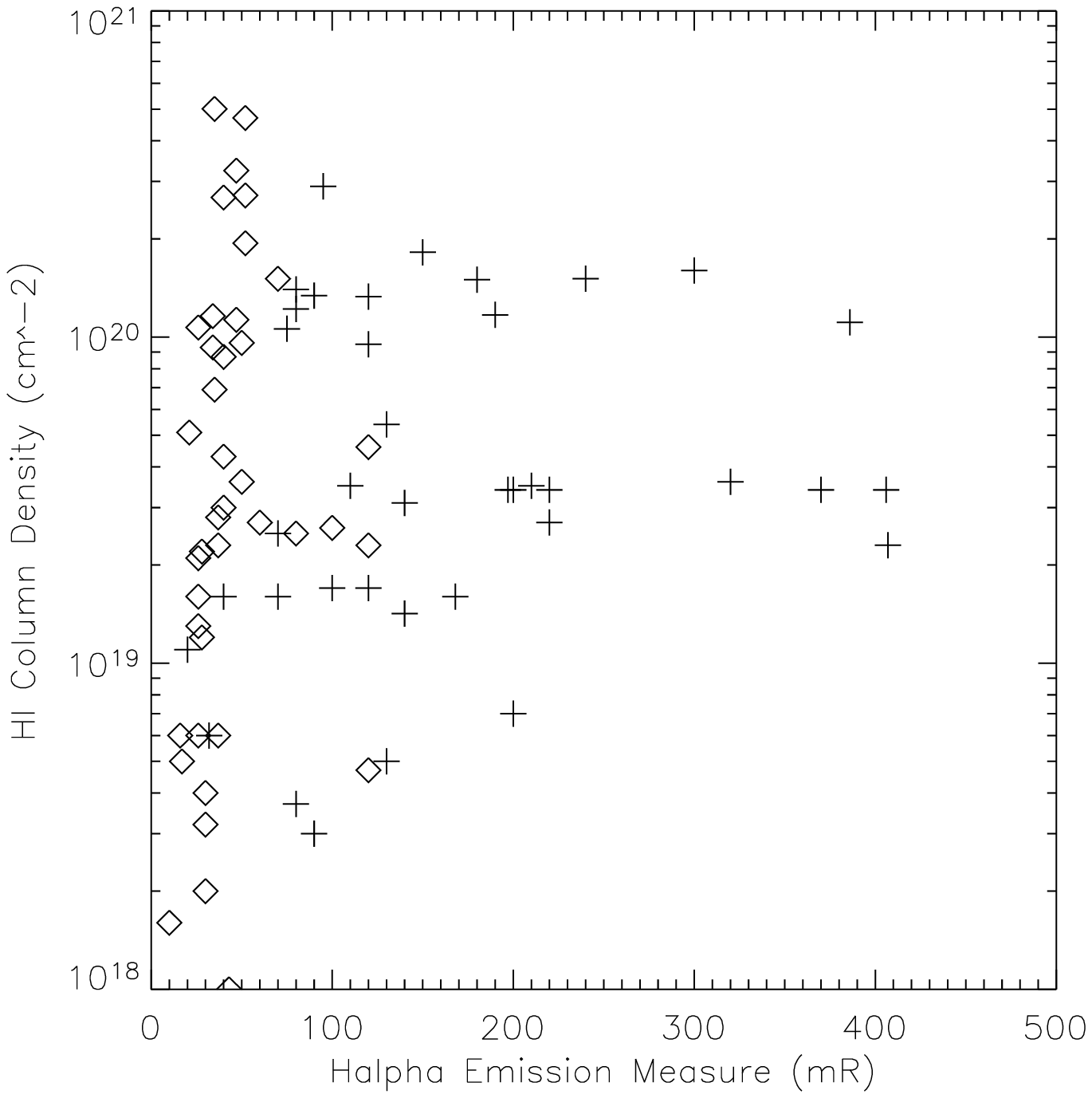,width=10cm,angle=0}}

\noindent {\sl Fig.~3.}
The relationship between the \Ha\ emission measure and the \HI\ column density for the
same data shown in Figure 2.  Detections are represented by crosses and non-detections
are represented by open diamonds.  The \Ha\ emission measures are NOT extinction corrected.
Using the extinction corrected values does not greatly change this plot, as can
be noted from the values listed in Table 1 and 2.

\medskip

\centerline{\psfig{figure=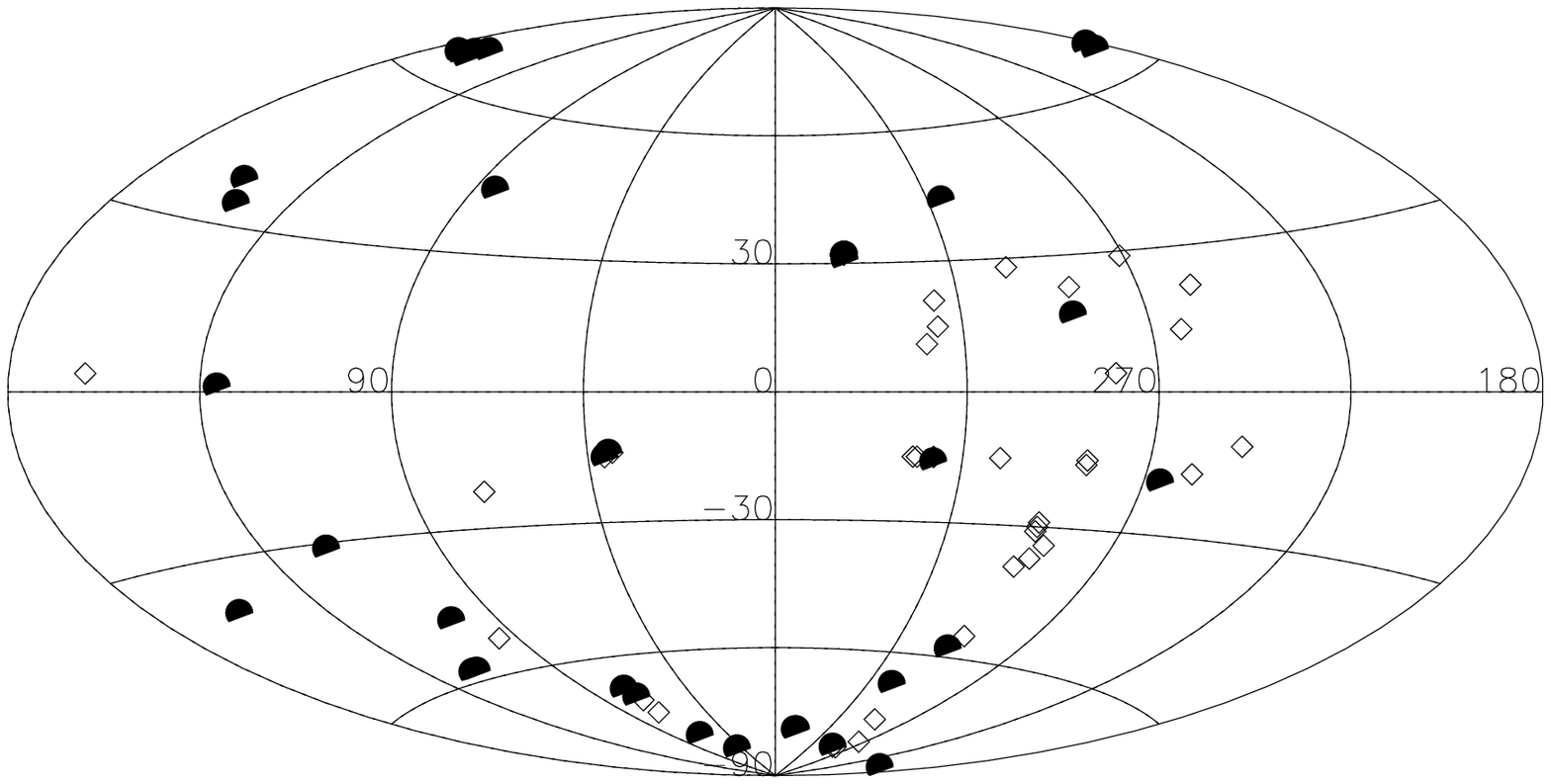,width=15cm,angle=90}}
\noindent {\sl Fig.~4.}
The distribution of \Ha\ detections (solid symbols) and non-detections (open diamonds)
of the same data shown in Fig. 2 in Galactic coordinates.

\medskip

\centerline{\psfig{figure=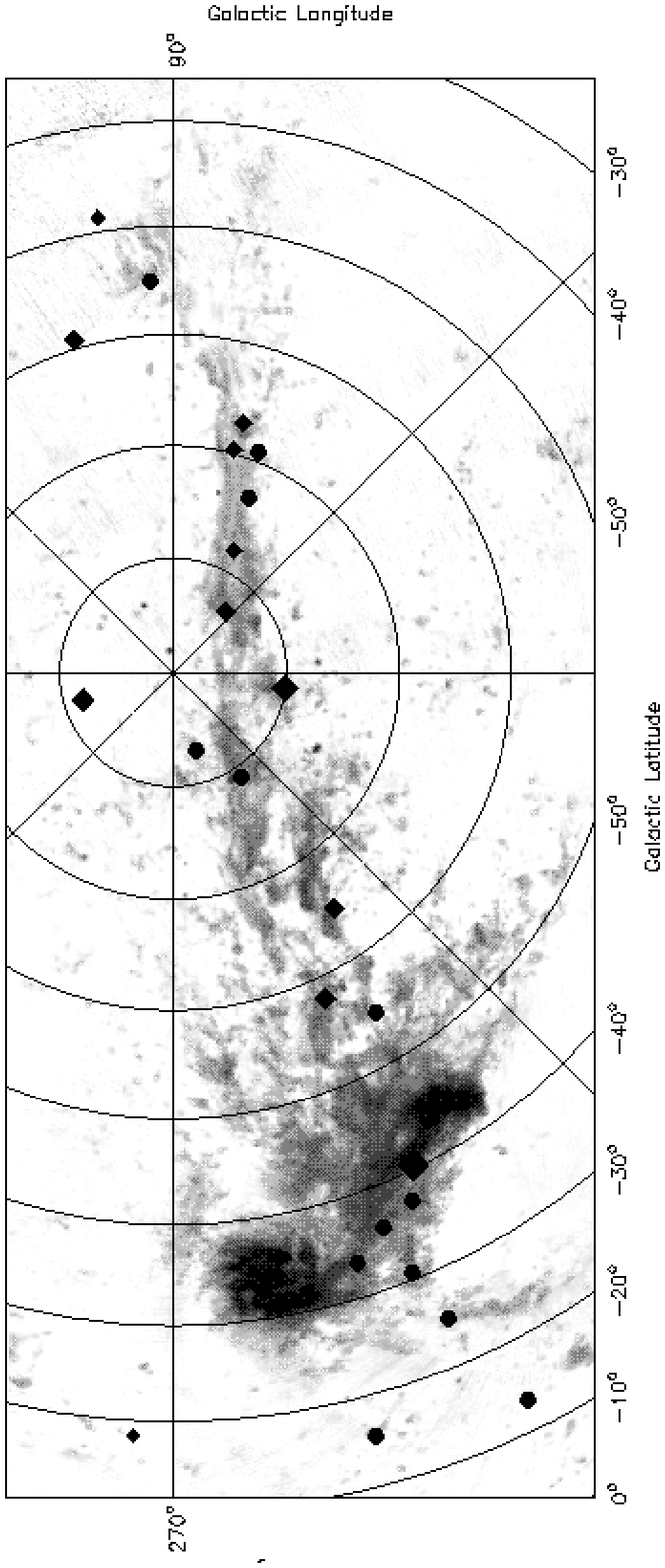,width=9cm,angle=0}}
\noindent {\sl Fig.~5.}
An \HI\ map of the Magellanic System showing column densities greater
than $2 \times 10^{18}$ cm$^{-2}$ (P03), with the
\Ha\ detections and non-detections labeled as diamonds and circles, respectively.
The size of the diamond represents the relative strength of the \Ha\ detection.
The positions and strengths of the \Ha\ observations were labeled by eye, and
are for general reference only.  The detections include this work, the WW96 observations, and two of the
CHVCs detected by T02 which are located near the northern tip
of the Magellanic Stream.
\medskip

\centerline{\psfig{figure=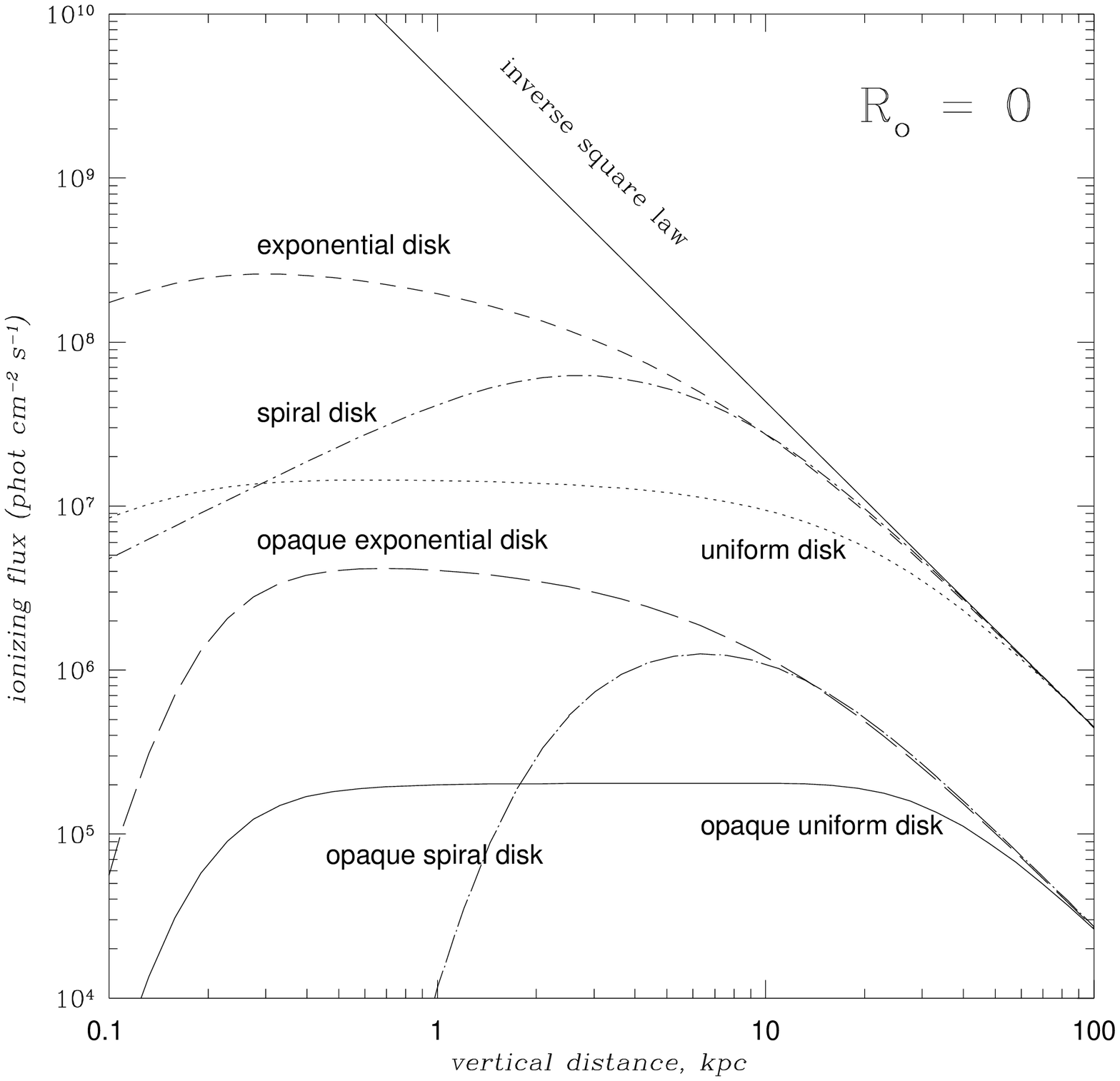,width=10cm,angle=0}}
\noindent{\sl Fig.~6.} The halo ionizing flux for different disk distributions 
(uniform emissivity, exponential and spiral)
compared to a simple inverse square law. 
The vertical distance is measured from the center of the disk along the 
polar axis. The top three curves are in the absence of dust and converge 
in the far field limit. The lower three curves include the effects of
dust where \tauLL $=$ 2.8 (\fesc\ = 6\%).

\centerline{\psfig{figure=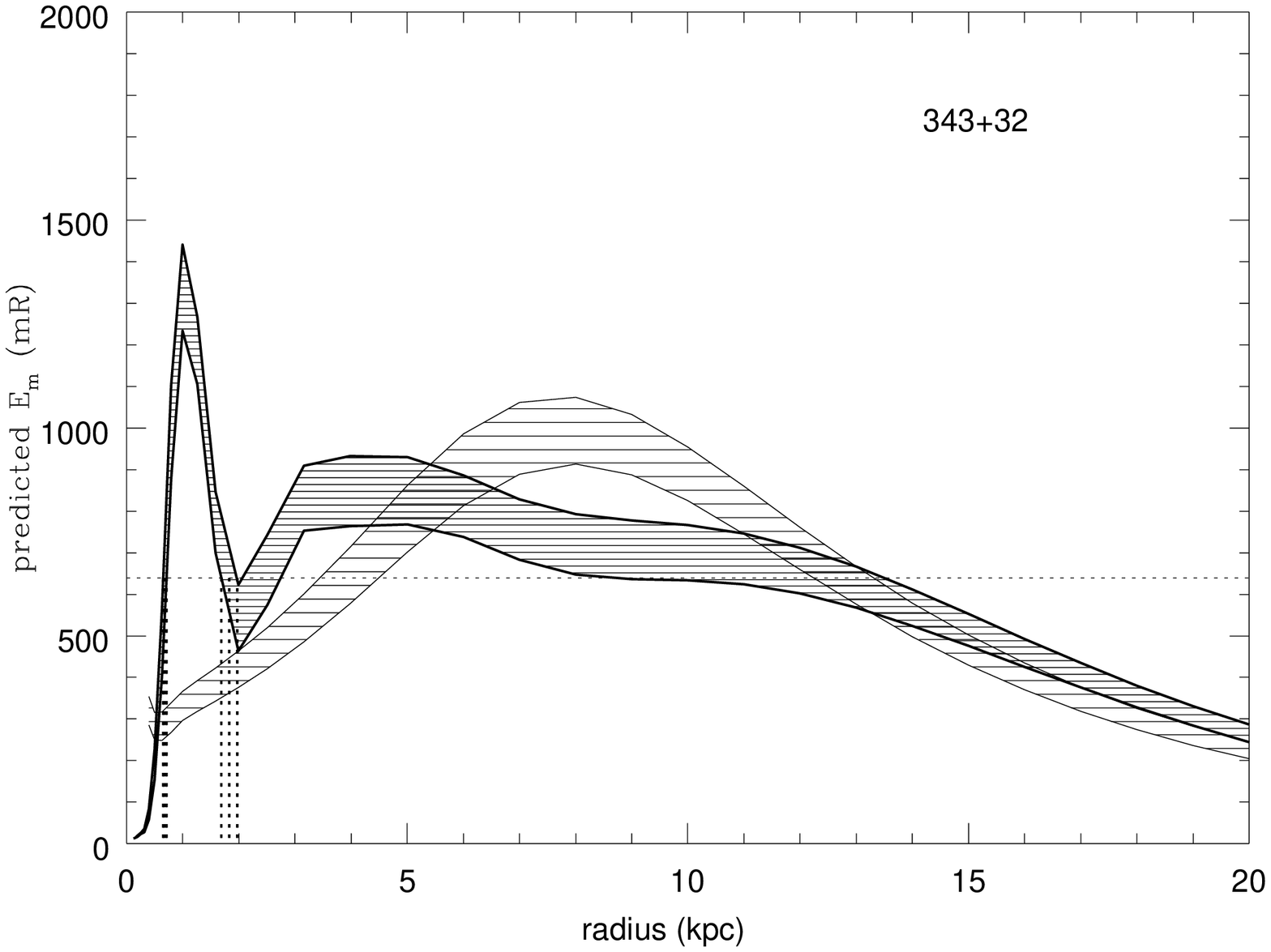,width=10cm,angle=0}}

\noindent {\sl Fig.~7.} Predicted run of emission measure (in mR) as a
function of radius (in kpc) along our sight line to Complex L.  The
light shaded model is the \Ha\ signal due to an exponential disk of
ionizing sources; the dark shading is for the spiral arm model. The
horizontal line shows our brightest observed E$_m$ for Complex L.  Note
the spiral arm model can produce multiple solutions depending
on the location of the HVC above the Galaxy. [We note that Weiner\etal\ (2001) 
detect emission measures of $\sim$1~R for a different cloud in Complex L,
indicating the near field solution (above a spiral arm) is correct.] Plots of
the model predictions for the other detected HVCs can be found at ftp://www.aao.gov.au/pub/local/jbh/disk\_halo.
\medskip

\centerline{\psfig{figure=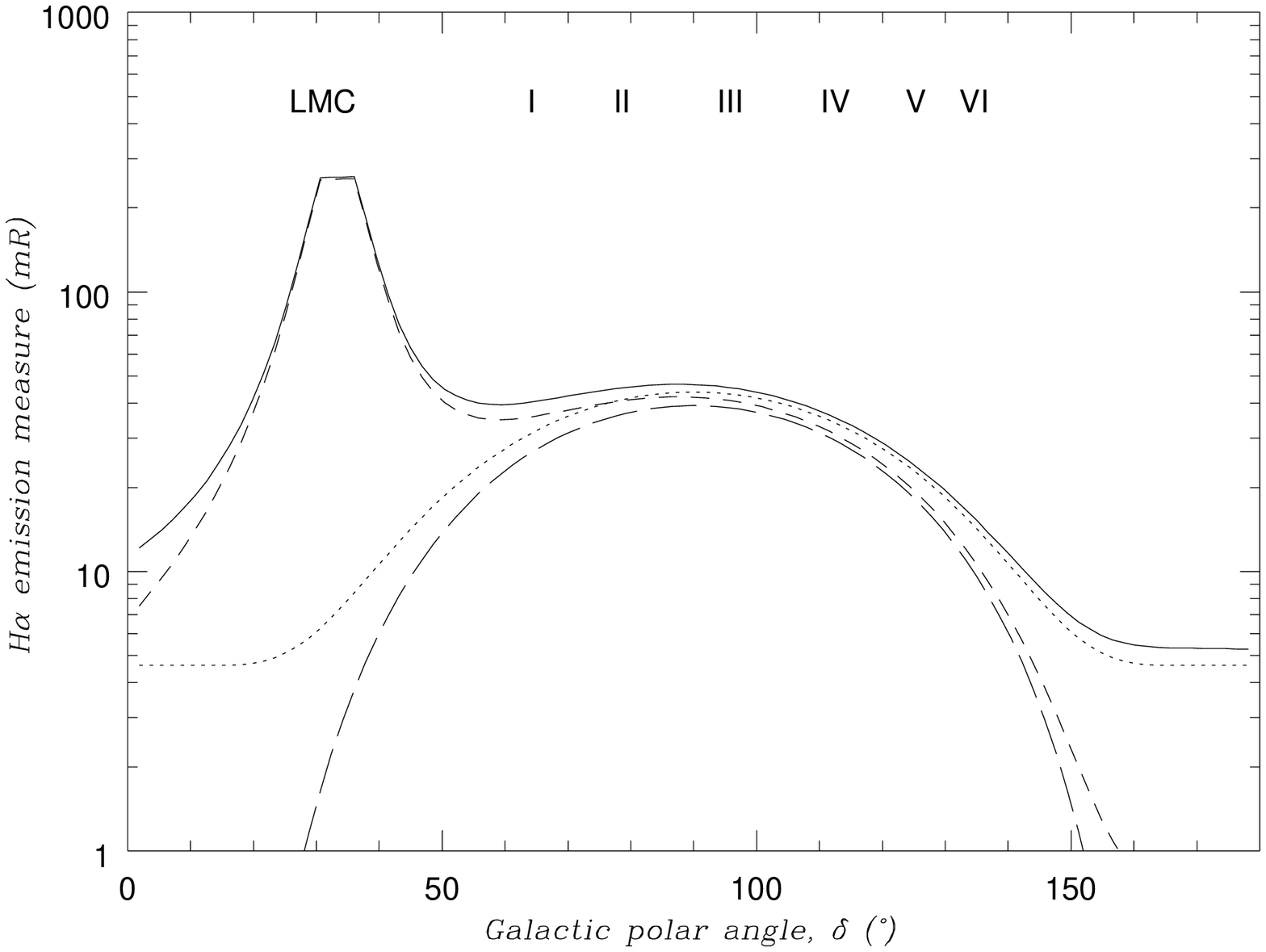,width=10.0cm,angle=0}}

\noindent {\sl Fig.~8} The predicted \Ha\ emission measure along the Stream as a
function of polar angle $\delta$ in units of log(mR) where $\delta=90$\deg\
is the South Galactic Pole.  The roman numerals refer to the specific
Magellanic Stream complex defined by Mathewson et al. (1977).  See
P03 for the definitions of these complexes on the map
shown in Figure 5.  The short dashed curve includes the
contribution of the LMC; the dotted curve includes the contribution of
a UV-bright stellar bulge. The solid curve includes the effect of the
LMC and a stellar bulge.
\medskip

\clearpage

\begin{center}
\scriptsize
\centerline{\sc Table of \Ha\ Emission Line Results, HI Properties, and Distances to Detected HVCs}
\smallskip
\vglue -0.12cm
\label{tab:res}
\begin{tabular}{llcccllccc}
\tableline
\multicolumn{1}{c}{ Obs.} & 
\multicolumn{1}{c}{ Common$^{a}$}  & 
\multicolumn{1}{c}{ $N_{HI}$} &  
\multicolumn{1}{c}{ $V_{lsr}$} &  
\multicolumn{1}{c}{ $\Delta$V$^b$} &   
\multicolumn{1}{c}{ E$_m$$^c$} & 
\multicolumn{1}{c}{ E$_{m(obs)}$$^d$ } & 
\multicolumn{1}{c}{ [NII]/\Ha\ } &
\multicolumn{1}{c}{ $V_{lsr}$} &
\multicolumn{1}{c}{ D$_{mod}$$^e$} \\
\multicolumn{1}{c}{ $\ell$ $b$} & 
\multicolumn{1}{c}{ Name} & 
\multicolumn{1}{c}{ ($10^{19}$ cm$^{-2}$)} & 
\multicolumn{1}{c}{ (H{$\rm\scriptstyle I$})} & 
\multicolumn{1}{c}{ km s$^{-1}$} & 
\multicolumn{1}{c}{ (mR)} & 
\multicolumn{1}{c}{ (mR)} & 
\multicolumn{1}{c}{ } & 
\multicolumn{1}{c}{ (${\rm H}\alpha$)} &
\multicolumn{1}{c}{ (kpc)} \\
\tableline		   	    		   	  
295.1$-$57.8 &  MS I (Fairall~9) &  9.5 &   191 &  52 &  128(D) & 120$^*$ & $<$ 0.25 & 200 &  0.5 - 25.7 \\
304.0$-$68.3 &  MS Ib  &  29.0     &    81     &  35    &  99  & 95  & $<$ 0.30 &   93 & 0.5 - 33.2\\
342.6$-$79.6 &  MS IIa  &    11.1       &  -120     &  37    &  407 &  386  & 0.15 & -124 & 1.7 - 9.7\\
342.2$-$79.9 &  MS IIa  &  3.4  &  -116  &  34  & 228(D) & 220 & $<$ 0.18 & -116 & 0.8 - 19.9\\
297.5$-$42.5 &  Bridge M  &  98.3     &   166     &  66    &  3796 & 3240 & 0.05 & 146 & - \\
040.3$-$15.1 &  Smith2$^f$    &  16.0       &    86    &  38    & 450 & 300 & 0.60  & 80 & 1.2 - 12.7 \\
040.6$-$15.5 &  Smith1$^f$    &    15.1      &    94    &  47    & 360 & 240 & 0.60 &100 & 1.2 - 13.4 \\
130.8+00.9 &   Complex H$^{g}$  &  18.2      &  -200   &  16    &  3697 & 150 & - & -170 & -   \\ 
170.9+64.7 &  Complex M W6  &    -      &    -    &  -     &  150  & 140$^*$ & -  & -90 & 1.7 - 9.6  \\
163.3+66.7 &  Complex M W2 &  11.7      &  -101   &  43    &  203  & 190$^*$ & -   & -90 & 2.2 - 6.7  \\
341.8+31.3 &  Complex L2$^h$      &   1.6   &  -146     &  58   &  263 & 168 & 2.5 & -124 & 0.5 - 19.9 \\
343.2+32.1 & Complex L3   &   3.6       &  -136     &  36    &  499 & 320 & 2.7 & -129 & 0.6 - 15.2\\
343.1+32.0 & Complex L4    &   3.4      &  -142     &  41    &  309 & 197 & 2.5  &-146 & 0.6 - 19.0\\
343.2+31.9 & Complex L5      &   3.4       &  -145     &  39    &  637 & 406 & 2.7 &-140 & 0.7 - 11.2\\
343.4+32.0 & Complex L6       &   2.3      &  -138     &  35    &  639 & 407 & 2.7 &-140 & 0.7 - 11.1\\
153.6+38.2 & Complex A$^i$ & 1.3 & -177 & 23 & 108(W) & 90 & - & -178 & 1.6 - 5.0 \\
084.3+43.7 & Complex C$^i$ & 0.54 & 120 & 15 & 133(W) &  130 & - & -111 & 1.9 - 14.2 \\
310.9+44.4 & HVC310.5+44.2+187 &  0.37  &  187 &  40 &  99(D)& 80$^*$ & 1.3 & 187 & 0.4 - 27.5\\
322.0$-$15.8 &  HVC321.7-16.0+113 &  1.7  &  113 & 59 & 125(D) &  100 & $<$ 0.50 & 113 & 0.5 - 18.5\\
104.2$-$48.0 & :HVC104.2-48-168$^i$ & 0.6 & -170 & 25 & 39(W) & 32 & - & -168 & 1.1 - 27.8 \\
118.5$-$58.2 & CHVC118.2-58.1-373$^i$ & 3.1 & -374 & 28 & 152(W) & 140 & - & -369 & 1.9 - 10.6 \\
119.2$-$30.8 & CHVC119.2-31.1-384$^i$ & 1.1 & -386 & 20 & 24(W)& 20 & -& -382 & 1.3 - 13.2 \\
158.0$-$39.0 & CHVC157.7-39.3-287$^i$ & 0.5 & -284 & 27 & 147(W) & 130 & - & -290 & 1.7 - 4.3 \\
197.4$-$81.8 & CHVC197.0-81.8-184 &  2.7 &  -184  &  41 &  227(D) &  220 & $<$ 0.30 &  -180 & 1.4 - 12.9\\
266.0$-$18.7 & CHVC266.0-18.7+336 &  1.42 & 336 &  31 &  190(D)&  140$^*$ & $<$ 0.60 & 336 & 1.2 - 6.1\\
285.9+16.6 &  XHVC287.6+17.1+111$^j$ &  0.7 &  111 &  32 & 241(D) & 180 & - & 111 & 0.8 - 9.9\\
\tableline
\\
\end{tabular}
\vskip-1.0mm
\noindent

{\scriptsize $^a$\,MS refers to a Magellanic Stream complex (Mathewson et al. 
1977), Smith is also Complex GCP, many objects are named with their 
catalog name from P02.}
{\scriptsize $^b$\,$\Delta$V at FWHM of HI line.}
{\scriptsize $^c$\,The emission measure in milliRayleighs (mR) has D in parentheses if the result is from the DBS and W if the result is from WHAM. All values are extinction corrected.}
{\scriptsize $^d$\,E$_m$ before the extinction correction.  The characteristic
detection errors are 10 mR, unless noted with a $*$.  The $*$ indicates that the \Ha\ line is within 2\AA\ of a skyline and the errors are between 15 - 30 mR. }
{\scriptsize $^e$\,Modeled distance based on E$_m$, the HVC position and the 
model described in B02 (\fesc\ = 6\% normal to the disk).  There is a near and far field 
solution based on the location of the HVC over the spiral arms.  The error on the distance
is generally less than 0.5 kpc for the near field solutions and less than 4 kpc for
the far field solutions and this incorporates the difference in using E$_m$ or E$_{m(obs)}$. 
Exceptions where the errors on the far field solutions are $\sim$9 kpc include
:HVC104.2-48-168 and CHVC119.2-31.1-384. 
Plots of the model predictions and specific error values can be found at ftp://www.aao.gov.au/pub/local/jbh/disk\_halo.}
{\scriptsize $^f$\,Results published in B98.}
{\scriptsize $^g$\,Unable to model distance because of location in Galactic 
Plane.  The dust correction may not be applicable at such low latitudes.}
{\scriptsize $^h$\,Weighted average for Complex L is E$_{m(obs)}$=300 mR, [NII]/\Ha\ =2.7, $V_{lsr} =-140$.}
{\scriptsize $^i$\,Emission line results from T98 and T02.}  
{\scriptsize $^j$\,Velocity of this cloud places [NII]$\lambda$6583 right on a skyline.}

\end{center}

\clearpage

\begin{center}
\footnotesize
\centerline{\sc Table of \Ha\ Emission Limits and HI Properties of Undetected HVC Positions}
\smallskip
\vglue -0.12cm
\label{tab:resu}
\begin{tabular}{llcccllc}
\tableline
\multicolumn{1}{c}{ Obs.} & 
\multicolumn{1}{c}{ Common$^{a}$}  & 
\multicolumn{1}{c}{ $N_{HI}$} &  
\multicolumn{1}{c}{ $V_{lsr}$} &  
\multicolumn{1}{c}{ $\Delta$V$^b$} &   
\multicolumn{1}{c}{ E$_m$$^c$} & 
\multicolumn{1}{c}{ E$_m(obs)$$^d$ } & 
\multicolumn{1}{c}{ $V_{lsr}$} \\
\multicolumn{1}{c}{ $\ell$ $b$} & 
\multicolumn{1}{c}{ Name} & 
\multicolumn{1}{c}{ ($10^{19}$ cm$^{-2}$)} & 
\multicolumn{1}{c}{ (H{$\rm\scriptstyle I$})} & 
\multicolumn{1}{c}{ km s$^{-1}$} & 
\multicolumn{1}{c}{ (mR)} & 
\multicolumn{1}{c}{ (mR)} & 
\multicolumn{1}{c}{ (${\rm H}\alpha$)}  \\
\tableline

293.4$-$56.4 & MS I  & 3.6, 9.6 & 226, 158 & 23, 40 & $<59$, $<54$ & $<55^*$, $<50$ & 226, 158 \\ 
342.2$-$79.9 &  MS IIa &    3.4      &  -116     &  34    &  $<52$[OIII] &  $<52$[OIII]  &  -116  \\
096.5$-$53.9 & MS V  & 4.6 & -366 & 45 & $<153$ & $<120^*$ & -366 \\
292.4$-$40.1 & Bridge1 & 50.1 & 184 & 53 & $<$ 42 & $<35^*$ & 184  \\
290.2$-$37.6 & Bridge2 & 47.0 & 198 & 74 & $<$ 79 & $<$ 52$^*$ & 198  \\
287.7$-$34.8 & Bridge3 & 26.8 & 204 & 41 & $<$ 52 & $<$ 40$^*$ & 204  \\
291.7$-$32.0 & Lead Arm1 & 27.2 & 222 & 33 & $<$ 71 & $<$ 52$^*$ & 222   \\
291.7$-$30.6 & Lead Arm2 & 19.4 & 236 & 43 & $<$ 78 & $<$ 52$^*$ & 236  \\
292.1$-$29.7 & Lead Arm3 & 5.1 & 305 & 37 & $<$ 30 & $<$ 21$^*$ & 305   \\
287.5+23.0 & Lead Arm4 & 11.3 & 238 & 36 & $<$ 70 & $<$ 47$^*$ & 238 \\
342.5+31.9 &  Complex L1$^{e}$ &  0.1 &  -126 &  20 & $<$ 65 & $<$ 43  &  -126  \\
248.5$-$12.2 & Pop EP1 & 0.6 & 334 & 58 & $<56$ & $<$ 16$^*$ & 334  \\
262.7+13.5 & Pop EP2 & 1.6 & 160 & 38 & $<41$ & $<$ 26 & 160  \\
280.1+04.0 & Pop EP3 & 6.9 & 163 & 35 & $<128$ & $<$ 35 & 163  \\
271.2+29.4 & Pop EP4$^{e}$ & 0.2 & 184 & 29 & $<39$ & $<$ 30$^*$ & 184  \\
326.5$-$14.6 & HIPASS J1712-64 &  0.4 & 458 & 41 &  $<44$ & $<$ 30 & 458   \\
040.6$-$15.5 &  Smith1 &   15.1 & 94 &  47 & $<$ 70[OIII] & $<$ 70[OIII] & 94   \\
039.3$-$13.8 & HVC039.3-13.8-233 & 3.1 & -233 & 29 & $<$ 213(D) & $<$120& -233  \\  
259.2$-$17.2 & HVC259.1-17.2+362 & 0.4 & 362 & 37 &  $<171$(D) &  $< 120$ & 362   \\
301.2+27.7 & HVC301.1+27.6+168 & 2.2 & 166 & 34 & $<37$ & $<$ 28 & 166  \\
321.5+20.7 & HVC321.7+20.8+167 & 3.0 & 166 & 36 & $<53$ & $<$ 40 & 166  \\
257.2+22.0 & :HVC257.2+21.9+188 & 3.1 & 188 & 33 & $<$ 93(D) & $<$ 80$^*$  & 188  \\ 
324.4+10.6 & :HVC324.4+10.6+151 & 4.3 & 151 & 47 & $<$ 167(D) & $<$ 100 & 151  \\  
162.0+02.5 & CHVC161.6+02.7-186 & 1.2 & -180 & 28 & $<$ 980 & $<$ 28 & -180 \\
284.6$-$16.1 & CHVC284.9-16.1+205$^{e,f}$ & 11.6 & 192 & 33 & $<48$ & $<$ 34$^*$ & 192  \\ 
285.6$-$83.3 & CHVC286.3-83.5+091$^{g,h}$  & 1.3, 0.6 & 86, -144 & 35, 44 & $<$ 26, $<$ 37 & $<$ 26, $<$ 37 & 86, -144 \\
289.7$-$83.0 & CHVC290.6-82.8+095$^{g,h}$ &  2.1, 1.6 & 95, -147 & 43, 37 & $<$ 26, $<$ 70 & $<$ 26, $<$ 70 & 95, -147 \\
306.3$-$16.0 & CHVC305.9-16.1+185 & 2.3 & 183 & 38 & $<61$ & $<$ 37$^*$ & 183  \\
321.0+14.9 & CHVC321.1+14.8+113 & 8.7 & 110 & 32 & $<73$  & $<$ 17 & 110  \\
275.2$-$80.7 & XHVC275.5-80.8-132 & 10.7 &  -139 &  82 & $<$ 26 & $<$ 26 &  -139  \\
290.9$-$76.3 & XHVC294.2-76.1+134$^{h}$ &  2.8, 0.03 & 141, -154 & 43, 15 & $<$ 37, $<$ 30 & $<$ 37, $<$ 30 & 141, -154 \\

\tableline
\\
\end{tabular}
\vskip-1.0mm
\noindent
{\scriptsize $^a$\,MS refers to a Magellanic Stream complex (Mathewson et al. 
1977), Smith is also Complex GCP, most objects are named with their 
catalog name from P02.}
{\scriptsize $^b$\,$\Delta$V at FWHM of HI line.}
{\scriptsize $^c$\,The emission measure limit in milliRayleighs (mR) has D in parentheses if the result is from the DBS and [OIII] if it is a limit on
the [OIII]$\lambda$5007 emission ($2\sigma$). All \Ha\ limits are extinction corrected.}
{\scriptsize $^d$\,E$_m$ before the extinction correction.   The characteristic
detection errors are 10 mR, unless noted with a $*$.  The $*$ indicates the \Ha\ line is within 2\AA\ of a skyline and the errors are between 15 - 30 mR.}
{\scriptsize $^e$\,Compromised by Fraunhofer lines from strong moonlight.}
{\scriptsize $^f$\, Also undetected by the DBS.}
{\scriptsize $^g$\, These clouds have been associated with the Sculptor dSph by Carignan et al. (1998).}
{\scriptsize $^h$\, There is also a negative velocity cloud, XHVC288.4-81.8-109, along this sightline.}
\end{center}

\end{document}